**Human-LLM Collaborative Construction of a Cantonese Emotion Lexicon**

Yusong Zhang[1], Dong Dong[1*], Chi-tim Hung[1], Léonard Heyerdahl[2], Tamara Giles-Vernick[2], Eng-kiong Yeoh[1]

1. Centre for Health Systems and Policy Research, Jockey Club School of Public Health and Primary Care, The Chinese University of Hong Kong, New Territories, Hong Kong
2. Anthropology & Ecology of Disease Emergence Unit, Department of Global Health, Institut Pasteur/Université Paris Cité, Paris, France,

*Corresponding author: Dong Dong, 4/F, School of Public Health Building, Prince of Wales Hospital, Shatin, New Territories, Hong Kong

Email: dongdong@cuhk.edu.hk

**Abstract**

Large Language Models (LLMs) have demonstrated remarkable capabilities in language understanding and generation. Advanced utilization of the knowledge embedded in LLMs for automated annotation has consistently been explored. This study proposed to develop an emotion lexicon for Cantonese, a low-resource language, through collaborative efforts between LLM and human annotators. By integrating emotion labels provided by LLM and human annotators, the study leveraged existing linguistic resources including lexicons in other languages and local forums to construct a Cantonese emotion lexicon enriched with colloquial expressions. The consistency of the proposed emotion lexicon in emotion extraction was assessed through modification and utilization of three distinct emotion text datasets. This study not only validates the efficacy of the constructed lexicon but also emphasizes that collaborative annotation between human and artificial intelligence can significantly enhance the quality of emotion labels, highlighting the potential of such partnerships in facilitating natural language processing tasks for low-resource languages.

**INTRODUCTION**

In recent years, with the emergence of large language models (LLMs) trained on extensive corpora of textual data, such as OpenAI's GPT series [1,2] and Google PaLM [3], artificial intelligence (AI) has climbed to new peaks in language understanding and generation [4,5]. Research on using LLMs to handle natural language processing (NLP) tasks has flourished, aiming to alleviate the limitations typically associated with traditional annotation methods that rely on human annotators, which are typically constrained by cost and scale [5,6]. Veselovsky et al. [7] discovered that crowdsourced workers on Amazon Mechanical Turk (MTurk) have been using ChatGPT to complete crowdsourcing tasks. Studies [8-10] have validated the cost-effectiveness and competitive potential of using LLMs for automatic annotation tasks. Li [11] compared the performance of crowdsourcing and LLMs in annotation quality based on carefully selected baseline datasets [12,13], highlighting the importance of introducing LLM annotators to improve the quality of crowdsourced aggregated labels. However, for specific domain tasks, LLMs may introduce biases compared to human annotations. Paradigms for human-LLM collaboration in annotation tasks have been explored. For example, Wang et al. [14] proposed a multi-step framework to enhance the cooperation in terms of trust and annotation performance by generating additional LLM explanations [15-17].

As a domain involving the determination of speakers' opinions, beliefs, feelings, and speculations regarding target entities [18], sentiment analysis can be conducted on diverse social media platforms to understand the public's emotions and attitudes towards various topics. A crucial component in emotion detection algorithms is the emotion lexicon, which consists of terms paired with their corresponding emotions [19]. Traditional emotion lexicons like WordNet Affect Lexicon [20] rely on high costs and significant efforts in expert labeling, making it difficult to scale up coverage. NRC EmoLex [19] carefully selected terms and annotated basic and prototypical emotions through Amazon's Mechanical Turk service, constructing a larger lexicon at a lower cost. The emotion labeling dimensions include anger, anticipation, disgust, fear, joy, sadness, surprise and trust [21]. Studies [22, 23] validated the effectiveness of ChatGPT in annotating sentiment text datasets, while Nasution and Onan [5] further examined the competitive performance of GPT-4 for low-resource languages.

As a Chinese variant with over 85.5 million speakers [24], Cantonese faces a scarcity of large-scale and high-quality data suitable for NLP applications, rendering it a low-resource language. This limitation is tied to its intrinsic characteristics of multilingualism and colloquialism, often resulting in mixed Chinese and English terms and terms derived from pronunciation, accompanied by weak grammatical rules [25-27]. Although numerous emotional lexicons and corpora have been developed for automatic sentiment analysis in Mandarin, Cantonese frequently incorporates vocabulary not found in Mandarin, leading to unsatisfactory performance of Mandarin emotion detection systems for Cantonese input [26]. Studies have been initiated to furnish linguistic resources for Cantonese emotion detection. OpenRice, a popular food review social platform in Hong Kong, represents a significant source of emotional Cantonese text data, as reviews and ratings reflect attitudes towards restaurants [28,29]. Local forums in Hong Kong also serve as resources for gauging public opinions on hot topics such as the Anti-Extradition Law Amendment Bill movement [30] and COVID-19 [31]. The disparities between spoken and written Cantonese expressions were further emphasized in [25], and a corpus of Cantonese verbal comments was constructed and analyzed.

The findings of [32] underscored the significance of lexical mapping in developing resources for language variants. Leveraging a parallel corpus of Cantonese speech from Hong Kong television broadcasts and Mandarin subtitles [33], a knowledge-based machine translation system was proposed based on a vocabulary mapping [34]. An additional Chinese corpus annotated for emotions on Weibo [26] was incorporated to construct an emotion detection system for Cantonese. In order to enhance the performance of a lexicon-based sentiment classifier, Ngai et al. [35] enriched the base Chinese lexicon with terms from comprehensive Cantonese lexicons. Considering the potential of LLMs for low-resource language annotation, this study takes advantage of human-LLM collaboration to construct an extended version of Cantonese emotion lexicon based on NRC EmoLex, combined with commonly used colloquial emotional terms extracted from Hong Kong local forums. Three textual emotion detection datasets originally in Chinese, English, and Cantonese were exploited to evaluate and validate the promising performance of our proposed lexicon.

**METHODS**

In this section, we initially introduced two tasks for constructing the Cantonese emotion lexicon, translation validation and emotion annotation. ChatGPT was prompted as a Cantonese expert for conducting emotion labeling. Human annotators were recruited and selected to carry out each of the two tasks. To facilitate data annotation and collection, an application based on Google AppSheet was developed.

**Task Description**

**Translation validation**: The Traditional Chinese version of the NRC EmoLex contains 14,154 words. We instructed annotators to verify the accuracy of the Chinese translation for each original English word and to identify any more idiomatic Cantonese expressions, and to supplement the translation if available.

**Emotion annotation**: On popular online forums in Hong Kong, more colloquial Cantonese expressions are commonly used in replies to thread topics, reflecting users' subjective opinions. We aimed to extract a broader range of authentic Cantonese vocabulary from these sources. Web crawlers were deployed to collect thread topics and their replies from two commonly used local forums, Discuss HK and HK Golden. The PyCantonese and Jieba Python packages were employed for text segmentation, with part-of-speech tagging recorded for the resulting terms. Term Frequency-Inverse Document Frequency (TF-IDF) is a metric used to assess the relevance of terms to the collected texts, calculated as follows:

$$TF(t,d) = \frac{\text{number of times } t \text{ appears in } d}{\text{total number of terms in } d}$$

$$IDF(t) = \log\left(\frac{\text{number of documents in corpus}}{1 + \text{number of documents containing } t}\right)$$

$$TF - IDF(t,d) = TF(t,d) \times IDF(t)$$

where $TF(t,d)$ is the term frequency of term $t$ within document $d$ and $IDF(t,D)$ is the inverse document frequency of term $t$ within document set, which is used to describe how common $t$ is in the corpus. The content concatenation of a thread topic and its replies represent a document. We retained the top 20,000 words in the TF-IDF score ranking.

**Annotation with ChatGPT**

Research has shown that excellent LLMs can serve as powerful tools to enhance the text annotation process, thereby improving the quality of aggregated labels [11]. We utilized ChatGPT, specifically the gpt-3.5-turbo model, to perform the task of translating English terms from the NRC EmoLex into colloquial Cantonese and annotating emotional terms extracted from Hong Kong forums. The emotion annotation maintained the dimensions consistent with the original NRC lexicon, encompassing anger, anticipation, disgust, fear, joy, sadness, surprise, trust, as well as the emotional polarities of negative and positive. We employed a prompt iteration strategy to determine the following prompts provided to ChatGPT.

> **Translation validation**: As a native Cantonese speaker living in the United States, you are teaching the local people to speak Cantonese. I'll give you a list of English words. Please translate them into colloquial Cantonese expressions, and finally output them in JSON format, where the key is the original English word, and the value is the Cantonese word. Each word should be as concise as possible.
> **Emotion annotation**: As an expert in understanding Cantonese texts, you can recognize Cantonese words with distinct emotions and describe them with the following basic emotions: anger, anticipation, disgust, fear, joy, negative, positive, sadness, surprise, trust. I will give you a vocabulary list, and some of these words are just neutral while some can have more than one type of emotions. Please identify the words that have very distinct and clear emotions, and output in JSON format, where the key is the word, and the value is the list of emotions that the word has.

For the former task, ChatGPT returned corresponding Cantonese vocabulary for the original words; for the latter task, ChatGPT provided us with 12,362 emotionally labeled words. Due to the influence of the training corpus, some proper nouns like names of celebrities and locations were also assigned emotion labels. We further filtered and retained 5,718 words with Chinese parts of speech tags including a (adjective), ad (adjective as adverbial), ag (adjectival morpheme), an (adjective with noun function), b (differentiating word), g (morpheme), h (preceding component), i (idiom), j (abbreviation), l (idiomatic expressions), q (quantifier), v (verb), vn (verb with noun function), and z (status word). These categories were observed to have a high proportion of emotion-related words. Half of this subset was randomly sampled for manual emotion annotation, accounting for 23.1% of the total AI-annotated vocabulary.

**Annotator Recruitment and Task Implementation**

We recruited native Cantonese speakers as volunteer annotators at the university to improve the Cantonese emotion lexicon. To facilitate data annotation and collection, we designed an APP using Google AppSheet for annotations to easily fill out the information on their phones, tablets, or computers.

For the translation validation task, each annotator was presented with an English word and its corresponding traditional Chinese version on each page of the app, as shown in Figure 1(a). If the annotator believed there was a colloquial Cantonese expression for the English word, they could input their answer in the "Your answer" field. For example, if "pretty" is translated as "漂亮", they can add an alternative translation such as "靚" and "正". If they agreed with the provided Chinese translation and had no further suggestions, they could proceed to the next word. We recruited a total of 51 annotators, divided into three groups: A, B, and C, each with 17 individuals. The dictionary comprised a total of 14,154 words, randomly divided into 17 portions, with the first 10 portions containing 833 words each and the remaining seven portions containing 832 words each. Each portion's word list was copied three times and distributed to the corresponding annotators in groups A, B, and C for cross-checking annotations using the AppSheet application.

Figure 1(a). An example of the APP interface for translation validation

Figure 1(b). An example of the APP interface for emotion annotation

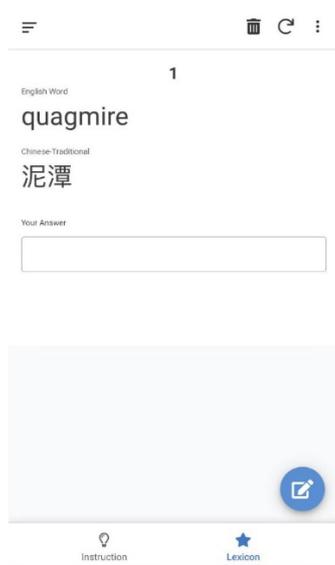
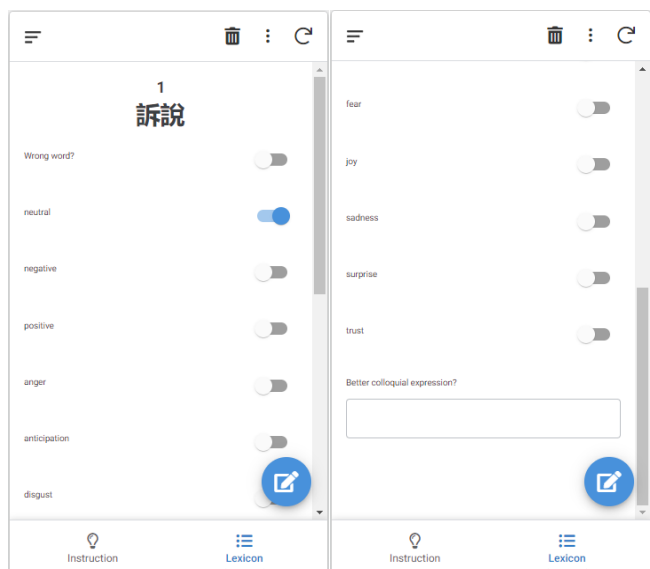

For the emotion annotation task, each annotator was presented with a Chinese word and all available emotional label dimensions on each page of the app, as shown in Figure 1(b). Annotator needed to select and tick the emotion labels they believed applied to the word. Additionally, two extra fields were provided: if the word did not exist, they could select "Wrong word," and if there was a better colloquial expression for the word, they could write it in the "Better colloquial expression" field. We initially created a demo to help select the three most consistent annotators for this task. Thirty words were randomly selected for candidates to provide preliminary annotation attempts. We used Krippendorff's Alpha on the 32 completed demo submissions to calculate the consistency among each trio and measure the extent to which the resulting data could reliably represent the real thing [36-38]. The three candidates with the highest consistency of 0.64 were selected for the subsequent task.

Considering that each word in English can correspond to multiple Cantonese expressions, we incorporated the Cantonese expressions supplemented by both human and LLM annotators in the translation validation task into the emotion lexicon. For the newly annotated words in the emotion annotation task, we determine the emotion labels for each word by majority vote.

**Consistency-Based Evaluation of Emotion Extraction Effectiveness**

Due to the lack of a benchmark dataset for validating the accuracy of the Cantonese emotion lexicon, we proposed a validation approach based on consistency. Inspired by the three resources utilized to construct the Cantonese emotion detection system in [26], three emotion text datasets with different original languages were collected to evaluate the performance of the lexicon in emotion extraction tasks. The OpenRice dataset in [29] was utilized for Cantonese sentiment polarity analysis, categorizing sentiments using star ratings given by users. Given the limited availability of public Cantonese emotion datasets, we gathered two emotion text datasets originally in English and Mandarin. The Bing Translator API was employed to acquire corresponding versions in the other two languages for each dataset. Specifically, CARER [39] assembled a dataset of English tweets through the creation of a set of hashtags and annotated it via distant supervision, encompassing emotion labels such as sadness, joy, love, anger, fear, and surprise. Ren-CECps [40] is a Mandarin emotion corpus with more nuanced emotion labels, including expectancy, joy, love, surprise, anxiety, sorrow, angry, and hate. Keyword-matching emotion extraction was conducted on each language version of each dataset using the corresponding language lexicon. To further assess the usability and effectiveness of our finalized extended Cantonese emotion lexicon, the extraction results of the NRC EmoLex, the emotion lexicon with widely validated validity, in English texts were used as a baseline to measure the performance of a lexicon on this task, i.e., the agreement of the results with the NRC EmoLex for the generated lexicon.

## RESULTS

In the annotation process, we calculated the intercoder reliability and obtained Krippendorff's alpha values among three annotator groups for the translation validation task (0.651) and among three annotators for the emotion annotation task (0.606). Notably, in the emotion annotation task, two annotators collaborated with ChatGPT and achieved a reliability of 0.663, demonstrating the quality of collaboration between LLM and human annotators. Consequently, the annotations provided by these two human annotators and ChatGPT with the highest reliability coefficient were used to determine the emotion labels for each vocabulary.

We retained only non-neutral words, resulting in a final lexicon with 11,433 words called EmoLex-Cantonese-Collaboration. The proportion of each emotion in the lexicon is illustrated in Table 1. Remark that each word may have more than one emotion label, and the dimensions of positive and negative are selected by annotators in cases where the emotion label is uncertain but the sentiment polarity is clear. The proportion of vocabulary labeled as negative surpasses that of positive labels. And the top five emotion labels in rank order are fear (0.178), trust (0.151), anger (0.147), sadness (0.144), and disgust (0.129), indicating a noticeably higher frequency of negative emotions compared to positive ones. Among the 6451 words with emotions included in the NRC lexicon, 3091 (47.9%) were annotated with an additional Cantonese expression, 1045 (16.2%) were annotated with two additional expressions, and 302 (4.7%) were annotated with three or more additional expressions. Besides, a substantial 62.8% of the words were provided with different expressions by ChatGPT, while human annotators only provided additional expressions for 26.1% of the words.

Table 1. The proportions of emotion labels in EmoLex-Cantonese-Collaboration

| Emotion label | Proportion |
| --- | --- |
| negative | 0.481 |
| positive | 0.381 |
| anger | 0.147 |
| anticipation | 0.115 |
| disgust | 0.129 |
| fear | 0.178 |
| joy | 0.087 |
| sadness | 0.144 |
| surprise | 0.068 |
| trust | 0.151 |

Table 2. Cohen's Kappa coefficients with annotation results of EmoLex-English for each lexicon

| Lexicon \ Dataset | OpenRice | CARER | Ren-CECps |
| --- | --- | --- | --- |
| EmoLex-Chinese (Mandarin) | 0.779 | 0.319 | 0.383 |
| EmoLex-Chinese (Cantonese) | 0.735 | 0.275 | 0.360 |
| EmoLex-Cantonese-Human (Cantonese) | 0.821 | 0.319 | 0.399 |
| EmoLex-Cantonese-ChatGPT (Cantonese) | 0.807 | 0.370 | 0.411 |
| EmoLex-Cantonese-Collaboration (Cantonese) | 0.889 | 0.370 | 0.439 |

To analyze the performance of different components in the annotation process, we also constructed and evaluated lexicons solely annotated by ChatGPT (EmoLex-Cantonese-ChatGPT) and solely by human annotators (EmoLex-Cantonese-Human). The original NRC Emotion Lexicon and its Traditional Chinese version are respectively denoted as EmoLex-English and EmoLex-Chinese. Cohen's Kappa coefficient [41] was employed to assess the agreement between the emotion extraction results of EmoLex-English on the English version of the dataset and the results of each modified lexicon to be validated on the corresponding language versions of the same dataset. As Cantonese and Mandarin share some linguistic overlap, we also assessed the performance of EmoLex-Chinese on the Cantonese version of the dataset for emotion extraction. The results are detailed in Table 2, with the language version of the dataset used indicated in parentheses. It is evident that our proposed EmoLex-Cantonese-Collaboration has shown the best performance across all three datasets. On the OpenRice, CARER, and Ren-CECps datasets, EmoLex-Cantonese-Collaboration outperformed EmoLex-Chinese, which is composed of Mandarin vocabulary, by 14.1%, 16.0%, and 14.6%, respectively. Furthermore, compared to the manually annotated EmoLex-Cantonese-Human, the improvements were 7.6%, 16.0%, and 10.0% for the same datasets.

DISCUSSION

Achieving the highest Cohen's Kappa coefficients on all three datasets reflects the significant improvement brought about by EmoLex-Cantonese-Collaboration in emotion extraction based on a lexicon for Cantonese. In particular, the consistency between our lexicon and EmoLex-English reaches 0.89 on OpenRice which is originally Cantonese. Limited by the scarcity of existing Cantonese resources, datasets derived through online translation lack sufficient localization. However, the results obtained from the other two datasets, which are not originally in Cantonese, show an approximately 15% improvement over EmoLex-Chinese, similar to the performance on the OpenRice dataset. Moreover, we observed a decline in performance when directly applying the Traditional Chinese version of the NRC lexicon to the Cantonese version of datasets compared to its application on the corresponding Mandarin version. This reaffirms the assistance lexical mapping provides in developing resources for low-register languages [26, 32-34]. It is inappropriate to directly use annotated corpora of one language for handling corresponding NLP tasks in another language variant, even if there are considerable similarities between them.

Observing the results of human annotators and ChatGPT annotators separately reveals that LLMs are highly competitive in Cantonese emotion lexicon annotation tasks, aligning with findings that LLMs exhibit outstanding performance in sentiment analysis tasks for low-resource languages [5]. Through collaboration between human annotators and LLMs via label aggregation, the quality of annotations in EmoLex-Cantonese-Collaboration is further enhanced compared to EmoLex-Cantonese-Human and EmoLex-Cantonese-ChatGPT. This is consistent with conclusions in [11] that augmenting a crowdsourced dataset with labels generated by superior LLMs can elevate the aggregate label quality of the dataset, surpassing the quality of LLM labels themselves.

**Limitations**

Our study has some limitations that can provide insights for future research. Firstly, our annotation of the lexicon lacked an expert refinement phase. Involving expert raters could further ensure the generation of high-quality labels [9]. Due to the absence of public benchmark datasets for Cantonese annotation, our evaluation only ensures the improved performance of the proposed lexicon in extracting emotions contained in text compared to the existing excellent lexicon. Efforts should be directed towards efficient ways and verified corpus to measure the quality of Cantonese annotations. And precise publicly available lexical mapping tools can facilitate knowledge transfer between Mandarin and Cantonese. Furthermore, we only deployed one type of LLM as an annotator, and testing more superior LLMs would enhance the robustness of the experiments. Additionally, it is worth investigating the introduction of explanations generated by LLMs at the right moment, which can aid in human-AI cooperation in terms of trust and annotation performance [14].

**Conclusion**

This study leveraged the collaboration of a Large Language Model (LLM) and human annotators to construct a new emotion lexicon for the low-resource language Cantonese. By aggregating emotion labels provided by the LLM and human annotators, we accomplished localization and extension of the NRC emotion lexicon. Three emotion text datasets were transformed and utilized to assess the consistency between our proposed emotion lexicon on emotion extraction and the NRC emotion lexicon. We validated the effectiveness of the lexicon and supported the efficacy of collaborative annotation between humans and artificial intelligence for label quality improvement.

**Acknowledgements**

We would like to thank Hospital Authority and Department of Health, Hong Kong Government, for providing the data for this study, and HyperLab for developing the web crawler. The Centre for Health Systems and Policy Research funded by the Tung Foundation is acknowledged for the support throughout the conduct of this study. Notably, we sincerely thank all the volunteers who participated in the annotation of this study and the list of volunteers will be appended to the final version of the manuscript.

This research was supported by Health and Medical Research Fund (grant numbers COVID190105, COVID19F03). The funder of the study had no role in study design, data collection, data analysis, data interpretation, writing of the manuscript, or the decision to submit for publication. All authors had full access to all the data in the study and were responsible for the decision to submit the manuscript for publication.

**Data availability:** Data is available from Dr. Dong DONG (Email. dongdong@cuhk.edu.hk) on valid ground.

**Conflict of Interest**

The authors declared that there is no conflict of interest in this research study.